\documentclass[runningheads]{llncs}
\usepackage[T1]{fontenc}
\usepackage{graphicx}
\usepackage{hyperref}

\newcommand{\splitatcommas}[1]{%
  \begingroup
  \ifnum\mathcode`,="8000
  \else
    \begingroup\lccode`~=`, \lowercase{\endgroup
      \edef~{\mathchar\the\mathcode`, \penalty0 \noexpand\hspace{0pt plus 1em}}%
    }\mathcode`,="8000
  \fi
  #1%
  \endgroup
}

\begin{document}
\title{The role of conformity in opinion dynamics modelling with multiple social circles}
\titlerunning{The role of conformity in opinion dynamics modelling}
%
\author{Stanisław Stępień\inst{1}\orcidID{0000-0003-2401-256X}\and\\ 
Jarosław Jankowski\inst{2}\orcidID{0000-0002-3658-3039} \and
Piotr Bródka\inst{1}\orcidID{0000-0002-6474-0089} \and
Radosław Michalski\inst{1}\orcidID{0000-0002-0106-655X}}
\authorrunning{S. Stępień et al.}
%
\institute{Wrocław University of Science and Technology\\Department of Artificial Intelligence, 50-370 Wrocław, Poland\\
\email{229852@student.pwr.edu.pl \{piotr.brodka@pwr.edu.pl,radoslaw.michalski\}@pwr.edu.pl}\\
\and
West Pomeranian University of Technology\\Department of Computer Science and Information Technology \\70-310 Szczecin, Poland\\
\email{jjankowski@wi.zut.edu.pl}}
\maketitle              
\begin{abstract}

~\\
Interaction with others influences our opinions and behaviours. Our activities within various social circles lead to different opinions expressed in various situations, groups, and ways of communication. Earlier studies on agent-based modelling of conformism within networks were based on a single-layer approach. Contrary to that, in this work, we propose a model incorporating conformism in which a person can share different continuous opinions on different layers depending on the social circle. Afterwards, we extend the model with more components that are known to influence opinions, e.g. authority or openness to new views. These two models are then compared to show that only sole conformism leads to opinion convergence.

\end{abstract}

\section{Introduction}

We tend to believe that we have a single opinion on a given topic. Yet, one can say that we also manifest different behaviour and opinions when interacting with others from each social circle we are a part of. This is because our expressed opinions are a function of our honest worldview, the type of group that we are exposing it to, and multiple other factors~\cite{noelle1974spiral}. This leads to the existence of our multiple \textit{I}s, where each is slightly different from the other and dependent on local social pressures. This type of behaviour is not considered a symptom of any social disorder but is a natural slight adjustment of our expression to the setting we are currently in. Moreover, this does not apply to all of us, as some non-conformists always present their inherent opinion. Most people, however, typically employ some level of conformism or adaptation of opinions.

This work focuses on investigating which effects are observable when we model opinion dynamics in a setting in which we belong to multiple social circles but also expose conformism and other social phenomena, such as sociability, openness, or obeying authorities. In order to represent different social circles, we reached for the framework of multilayer networks that serve this purpose well. The main objective of this work is to demonstrate the outcomes of a simulation of a continuous opinion spread process in a setting in which we model multiple social circles as separate layers of a multilayer social network and also incorporate the aspect of conformism in such a way that individuals are capable of exposing different opinions at different layers. The agent-based simulations led to a better understanding of how continuous opinions fluctuate and whether this leads to differences in the opinions of individuals related to a variety of contexts they expose them. To make the model even more realistic, we also incorporate to it the influence of authority, openness to different views and sociability.

This work is organised as follows. In the next section, we present the related work, mostly underlining conformism, as it is the base of our model. Next, in Section~3, we present the conceptual framework, where we also refer to all other aspects the model incorporates (influence of authority, sociability and openness to new views). Section~4 is devoted to presenting two experimental scenarios: solely with conformism (Section 4.1) and with all the model components (Section 4.2). In Section~5, we draw conclusions and present future work directions.

\section{Related work}

Various aspects of human activity take place among others and the need for popularity or acceptance can be  one of the social targets. Social scientists explore how it leads to conformity due to the fact that not following accepted norms or standards  can lead to penalization of individuals due to their altitudes \cite{bernheim1994theory}. From another perspective, conformism can be positioned within social learning strategies, and an approach to copying the majority is based on conformist bias ~\cite{kendal2018social}.  Conformist transmission takes a vital role in effective social learning strategies~\cite{muthukrishna2016and}. The selection of social learning strategy is based on cognitive abilities, social status, and cultural background.  Conformism can slow the diffusion of innovations occurring in the population with low frequency~\cite{grove2019evolving}. They can be lost in conditions with conformist bias. In terms of awareness and, for example, epidemics, a high fraction of conformists within the network can lower efforts for epidemics suppression because of the focus on defection and the impact of anti-vaccination altitudes~\cite{miyoshi2021flexible}. From the perspective of cascade behaviours, sequences of individual decision and information cascades conformism can be supported by rewarding conformity institutions or at the community level~\cite{hung2001information}.

To better understand the mechanisms of conformism, several attempts were made to model the impact of conformism on social influence and information spreading. For example, an algorithm for analysing and detecting influence and conformism within social networks was proposed in~\cite{li2011casino}. The authors focused on a context-aware approach, and the method is based on the extraction of topic-based sub-graphs, with each sub-graph assigned to specific topics. Each node can be characterised by different influence and conformism measures for various issues. Positive signs represent support, while harmful distrust and opposition. Influence and conformity indices are computed for each node. The empirical study was performed on Epinions and Twitter datasets. Influence within networks was calculated using conformism based on positive and negative relations. Another work emphasised the lack of conformity factors within influence maximisation algorithms~\cite{li2015conformity}. Typical methods focus on influence parameters but are conformity-unaware. The authors proposed a greedy approach with integrated influence and conformity for effective seed selection. The critical element is the conformity-aware cascade model and its extension to the context-aware model with contextual information used. The method is based on partitioning of the network into sub-networks and computations within components. Influence and conformity indices are computed using the earlier presented method~\cite{li2011casino}.

The need for integrating conformity parameters in spreading models was emphasised in the SIS model applied to cultural trait transmission~\cite{walters2013sis}. The conformist influence was added as a conformity function resulting in an additional weight to transmission rates. Conformity bias was represented by a sigmoidal shape covering the tendency to follow the majority. Conformism was also studied for the SIR model, and a set of infected neighbours is added to overall probability and is integrated with transmission rate~\cite{lu2019impacts}. Conformity rate was also integrated within the social opinion formation model for vaccination decision-making with a probability of converting nodes to a social opinion of their neighbours implemented~\cite{xia2013computational}. 

\section{Conceptual framework}
\label{sec:materialsandmethods}

\subsection{Multilayer networks}
Conformity as an act of matching or changing one behaviour to adjust it to others' beliefs, opinions, attitudes, etc.~\cite{cialdini2004social} is hard to study using a simple one-layer network as they cannot reflect the full complexity of different behaviours and environments we are in. Every day we immerse ourselves in different social circles like family, friends, coworkers, colleagues from the basketball team, online friends and so on. Towards each of those cycles, we might present different 'masks' to be accepted, look better, feel comfortable, etc. For example, we might present our true opinion to our family and friends while we only pretend to accept the opinion of our coworkers because we are afraid that our true opinion won't be accepted in this social circle.
On the other hand, people we trust, like family or friends, are more likely to influence us to truly change our beliefs than our colleagues from the basketball team.

\begin{figure}
 \centering
 \includegraphics[width=.4\textwidth]{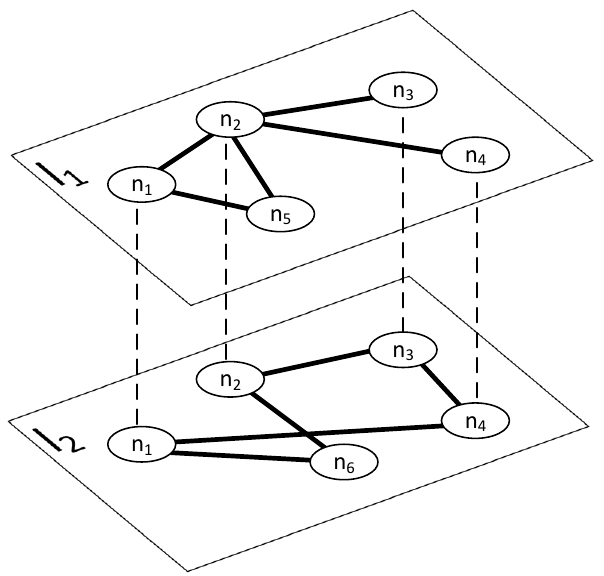}
 \caption{An example of a multilayer network}
 \label{fig:network}
\end{figure}

All these situations are next to impossible to model using a one-layer network, thus in this study, we have decided to use a multilayer networks~\cite{brodka2018multilayer,brodka2020interacting,dickison2016multilayer} which allows us to model different interactions and behaviour in various social circles. The multilayer network is defined as $M = (N,L,V,E)$~\cite{kivela2014multilayer}, where: 
\begin{itemize}
 \item $N$ is a not empty set of actors $\{n_1,..., n_n\}$,
 \item $L$ is a not empty set of layers $\{l_1,..., l_l\}$, 
 \item $V$ is a not empty set of nodes, $V \subseteq N \times L$,
 \item $E$ is a set of edges $(v_1, v_2): v_1, v_2 \in V$, and if $v_1=(n_1, l_1)$ and $v_2=(n_2, l_2) \in E$ then $l_1=l_2$.
\end{itemize}
The example of a multilayer network is presented in Figure~\ref{fig:network}. This network contains: six actors $\{n_1, n_2, n_3, n_4, n_5, n_6\}$, two layers $\{l_1,l_2\}$, ten nodes $\{v_1=(n_1,l_1), v_2=(n_2,l_1), v_3=(n_3,l_1), v_4=(n_4,l_1), v_5=(n_5,l_1), v_6=(n_1,l_2), v_7=(n_2,l_2), v_8=(n_3,l_2), v_9=(n_4,l_2), v_{10}=(n_6,l_2)\}$, and ten edges $\splitatcommas{\{(v_1, v_2), (v_1, v_5), (v_2, v_5), (v_2, v_3), (v_2, v_4), (v_6, v_9), (v_6, v_{10}), (v_7, v_8), (v_7, v_{10}), (v_8, v_9)\}}$.
To align with the naming convention used in related works, we often use the term \textit{agent} instead of \textit{actor}. However, both terms' meaning is the same in our paper.
 
\subsection{Continuous opinions}

Most opinion dynamics models derived from physics related to spin models treat opinion assigned to nodes in a discrete space with a binary representation~\cite{galam2002minority}. It is adequate when selecting only one of the options, e.g., \textit{pro} or \textit{contra}, \textit{convinced} or \textit{not convinced}. However, in reality, it is also possible to express attitudes toward specific situations to some extent. Growth of confidence can be modelled in a continuous process taking into account gradual changes with early models presented in~\cite{deffuant2000mixing,hegselmann2002opinion}. They assume that opinion is adjusted together with each contact and the ability to hear thoughts from others. In~\cite{hegselmann2002opinion} opinion of each agent evolves towards the average opinion of neighbouring agents.  Other mechanics are based on random pairwise encounters leading to possible compromise~\cite{deffuant2000mixing}. Continuous models were extended towards bounded confidence with interaction limited to agents with close opinions~\cite{lorenz2007continuous}. External opinions are taken into account only if the difference from the own opinion of the agent is lower than a certain bound of confidence. Other extensions include the ability to model heterogeneous confidence thresholds, disagreement, the wisdom of the crowd, or integrate a hybrid approach based on the degree assigned to discrete opinions~\cite{sirbu2017opinion}. Mechanics of continuous opinions in the proposed model and multilayer setup will be independent for each layer and interactions within it.  For each layer, the value of the agent’s opinion before and after the interaction with each agent will be stored and will be used for conformity measure computation.

\subsection{Authority}
Authority factors in how one may consider other individuals' credentials to be greater or lesser than these of ourselves~\cite{cialdini2004social}. Such a view of how knowledgeable another person may be as compared to us affects how likely we are to heed this person's influence~\cite{lewis1849essay}. To factor this into the model, a single parameter has been introduced, describing how authoritative a given agent is going to be when interacting with his or her neighbourhood. Real-world social influence is often dependent on how the influencing party is perceived by the influenced one. This is often the case when a person is interacting with a specialist or a professional in a given area. In such situations, one is much more prone to subduing or even completely reversing his or her personal view of the given matter if the person or people he or she spoke to are considered as of great authority. 
\par
This model's value of the parameter describing authority will be a floating point number ranging from 0.0 to 1.0, where 0.0 will signify a person without any authority whatsoever. This would refer to a real-life situation where if met with such a person, one would offhandedly disregard this person’s opinion on the matter as extremely uninformed and gullible. People tend to distrust or sometimes even ridicule information obtained from such individuals; we could revoke, for example, a situation where an inexperienced person (e.g. a child or a person whose opinion we consider to be outdated) tries to argue his or her opinion against a person who is much more acquainted with the matter of argument or holds a position of respect (e.g. a scientist or a courtroom judge). Of course, the degree to which one considers a person to be authoritative varies between individuals, and the one who could be considered to be of great knowledge and authority by Person A could just as well be considered to be of close-to-none authority by Person B. However, since the experiments will not focus on how one's authority changes, the model will not consider this factor, and each person will be given their own constant value of this parameter which will not later change throughout the course of the experiment.
\par 
During an interaction between two agents, the algorithm will have to take into account both the authorities of the two individuals in order to decide which one is considered to be more authoritative, as well as the average value of this parameter from one's neighbourhood, as to verify how a given individual compares with the rest of agents in one's neighbourhood~\cite{walton2010appeal}. During each interaction, a result of Equation~\ref{eq:authority_inequity} will be one of the factors deciding how this interaction will affect the agent's opinion.
\begin{equation}\label{eq:authority_inequity}
    A_{bl} < \bar{A}_{Nl}
\end{equation}
Where $A_{bl}$ denotes the value of agent B's opinion in the layer \textit{l} of the network and $\bar{A}_{Nl}$ is an average opinion of all of the agent A's neighbours in this layer. This check will determine whether the person who interacts with the given agent may be considered reliable and trustworthy or should rather be dismissed as unqualified depending on how close their opinion is to the agent. This may either make them consolidate more with the status quo of their neighbours or even remove them further away from it.


\subsection{Conformity}
Depending on one’s inclinations toward conforming behaviours influences of other people differ in effectiveness. This behaviour is predicated on multiple factors, such as one’s tendency to be subservient, amicable or simply obedient. One may distinguish two main types of influences which cause people to behave in a more conforming way: normative influence for situations when an individual would like to fit in with his or her group and informational influence for situations when one believes the group is better informed than they are~\cite{asch1940conformityExperiment}. 
\par 
As it is described in Asch's experiment, one is likely to alter its opinion on a given matter if other people around express different views, even if the individual knows this information is not true. To factor these into the model, each agent will be assigned a random float value from 0.0 to 1.0, where 0.0 denotes a person being practically immune to the influence of their environment and not afraid of displaying their own honest opinion regardless of how others view it. This could relate to an example of a very confident person who follows their chosen ideal exclusively because of its merit, not being affected by whether it is frowned upon by others. On the other end of the spectrum, a person with a value of 1.0 will most of the time conceal their true view on the matter and adjust the opinions that they express to others to what appears to be the most appropriate in the given situation. The rest of the spectrum will describe different degrees to which one is willing to alter a genuine opinion to adjust to the group.

\subsection{Openness to new views}
Having accounted for one's conformity when interacting with others and their opinions, the next step is to include the influence of one's cognitive bias. Even though there have not been too many models dealing with this problem, some noteworthy works have been created, like the one by Longzhao Liu's team~\cite{liu2021ConginitveBiasModel}, in which agents determine who belongs to their in-group or out-group based on the other neighbour's similarity in opinion and based on that a likeliness of interaction between the two is being determined.
\par
As a product of this bias, most individuals tend to downplay or downright disregard any data or opinions which disagree with their prior held beliefs~\cite{Zollo2017Tribalism} while acting in a more accepting way towards new ideas which are closer to their personal opinion. This problem is especially interesting in the context of a network of multiple layers in which one agent may already be altering his or her opinions to fit in more with a specific group. 
\par
The degree to which one is antagonistic or accepting towards views that stray from his or her own is predicated on a number of sociological and psychological factors, such as the specificity of one's community, gullibility, dogmatism, etc. We have decided to parameterise these factors with the use of a single coefficient which can be more broadly addressed as the \textit{openness} and assign each of the actors in the model with their own attribute, which value will be randomly set to a float from the scope of \begin{math}[0.0, 1.0]\end{math} where the value of zero would denote a person who is extremely close-minded and displays a very strong cognitive bias which causes her or him to ignore opinions which are even remotely different than their own and a value of one would indicate a person who does not discriminate even these opinions which are on the other end of the spectrum from their own.
\par
With this parameter, it will now be possible to include the influence of peer pressure in the model by comparing the difference between the opinions of two agents to the value of this parameter. Therefore an exemplary interaction between agent A and agent B will call out the formula as shown in Equation \ref{eq:comparing_diff_of_opinions_to_openness}. 
\begin{equation}\label{eq:comparing_diff_of_opinions_to_openness}
    \left | O_{bl} - O_{al} \right | < \bar{\theta }_{Nl}
\end{equation}
Where $O_{al}$ and $O_{bl}$ are consecutively agent's \textit{A} and agent's \textit{B} opinions in the layer \textit{l} of the network and $\bar{\theta}_{Nl}$ is the value of the average openness of agent's A neighbours. This equation may be interpreted as agent A assessing how acceptable a new opinion presented to them will be with their peers from the social circle where the interaction took place. The result of this inequality is going to be one of the two factors (next to the comparison of authority) deciding how this interaction is going to affect the agent's opinion.

\subsection{Sociability}
How often one interacts with others in different spheres of his or her life is a result of a variety of factors such as, e.g. one's character or social situation and background. It is not within the scope of this system to model these complex interactions and instead, we have decided to garner them into a single coefficient describing the likeliness of the given agent to form connections with the others and later interact with them.
\par
Similarly to the \textit{Openness} and \textit{Authority} parameters, \textit{Sociability} will also be assigned to be a floating point number from within the scope of [0.0, 1.0], where 0.0 will signify an individual who is extremely solitary and avoids all the interactions with their neighbours and 1.0 an agent who interacts with almost all of their neighbours in each step of the simulation.

\section{Simulations}
\subsection{Conformism in multiple social circles}
\subsubsection{Introduction}
\par 
This experiment is designed to illustrate the operation of the multilayer social network. It employs a relatively simple formula that does not use the other parameters and instead utilises only agents' conformity measures to decide how social influence affects one's opinions.
\par
This experiment consists of one thousand agents whose conformity and opinions have been initialised to be random floating point values from within their discussed scopes: \begin{math}[0.0, 1.0]\end{math} for Conformity and \begin{math}[-1.0, 1.0]\end{math} for each of the presented opinions in the layers. In this simulation the only factor deciding how one's opinion is influenced by the interaction with other actors is the agent's conformity parameter as described in the Formula \ref{eq:eq_opinion_change_exp1} ensures that conformity is a factor deciding to what degree a person will adopt others' opinion when exposed to it:
\begin{equation}\label{eq:eq_opinion_change_exp1}
 V({O}_{al},O_{bl})=(1 - C_{al}) \times O_{al} + C_{al} \times O_{bl}
\end{equation}
where \begin{math}
    O_{aln}
\end{math} denotes the new value of agent's \textit{a} opinion at the layer \textit{l} after the interaction with agent \textit{b} and similarly, \begin{math}
    O_{al}
\end{math} denotes value of agent's opinion before this interaction and \begin{math}
    O_{bl}
\end{math} stands for agent's \textit{b} opinion in the given layer.
\begin{math}
    C_{a}
\end{math}
denotes the value of agent's \textit{a} conformity measure.
\par
This formula entails that an agent whose conformity would be of a maximal value (equal to 1) would always fully adopt the opinion of a person he or she last spoke to. Similarly, an agent with the minimal value of this attribute would not alter his or her opinion no matter how many interactions took place.
\par
After each interaction, one's personal opinion is updated as well. The algorithm just assigns it to be an average of all the opinions displayed in the other layers of the simulation.

\subsubsection{Overview of the experiment}
\par
This experiment consists of a thousand steps of simulation with a thousand agents. Firstly, a multilayer network has been constructed. This one, similarly to others consists of five layers: 'personal worldview', 'workplace', 'personal life', 'social media 1' and 'social media 2'. Secondly, all of the experiment agents were initialised; this included assigning each individual with randomised values of the parameters, transferring their population into the network in the form of nodes and creating connections between the agents' nodes in each of the dimensions of the network. To obtain a relatively realistic number of particular agent's neighbours for each of the layers, the random parameter responsible for creating a connection between two individuals has been set to 0.05\% chance. A set of agent's neighbours, once set at the beginning of the experiment, is not going to be altered throughout its duration. After these initial conditions have been created, redetermined number of simulation steps is going to take place. In each step, for each of the layers, each of the agents may interact with his or her neighbours. Sets of one's neighbours may vary between the consecutive layers. Each of these interactions has a 5\% chance of taking place. Therefore in each of the steps, each agent may hold between 0 and \begin{math}
    l \times n
\end{math} interactions (\textit{l} being a number of layers and \textit{n} denoting one's number of neighbours in a given layer of the network).
\par

\subsubsection{Results of the experiment}
\begin{figure}
\includegraphics[width=\textwidth]{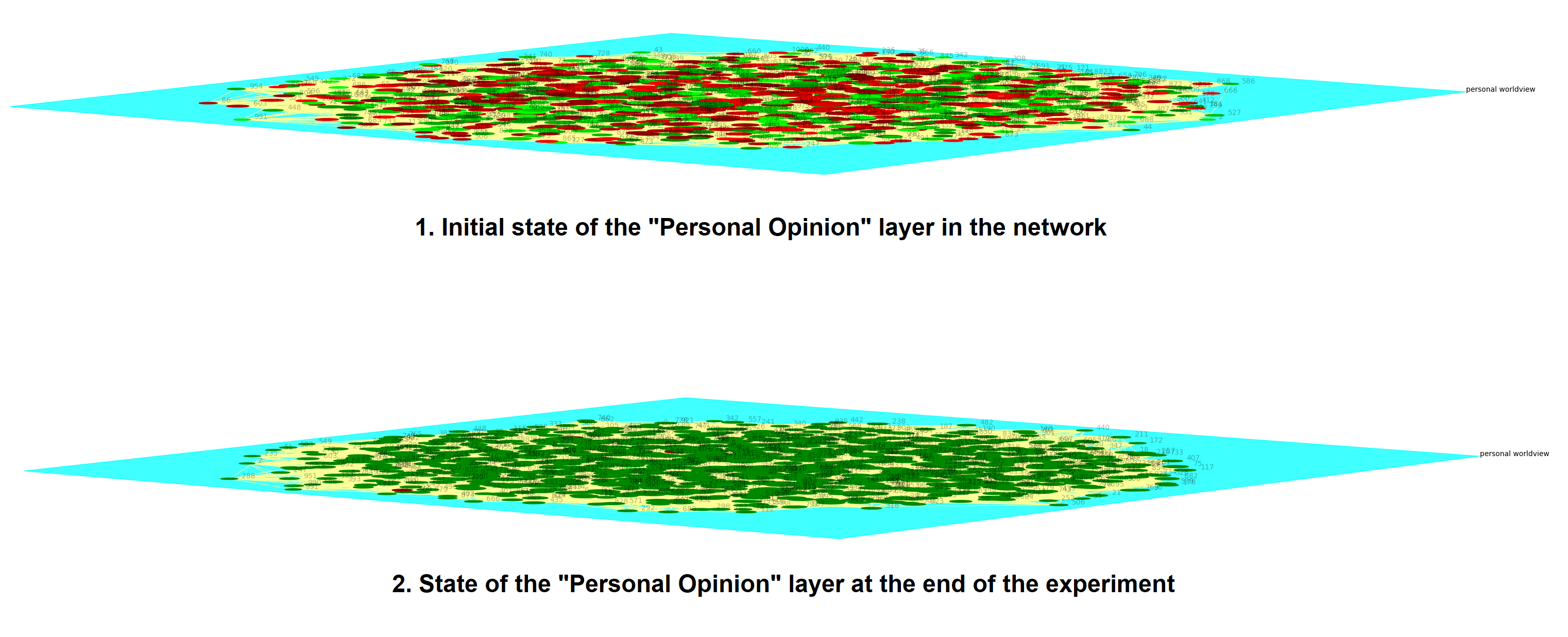}
\caption{Comparison of the state of the 'Personal Opinion' layer before (top of the figure) and after (bottom of the figure) the course of the experiment.} \label{fig:exp1_before_after}
\end{figure}

\par

The experiment illustrates how this version of the model operates and how the agents' opinions are being affected by the simplified social influence algorithm. As it is shown in Figure~\ref{fig:exp1_before_after}, the initial distribution of opinions has been quite random. However, throughout the run of the simulation, agents, as a group, have become more aligned in their personal opinions even in the remaining layers (which are not shown in this figure). This proves the fact that, as expected, agents' opinions seem to coincide more than before the experiment began.

\begin{figure}
\includegraphics[width=\textwidth]{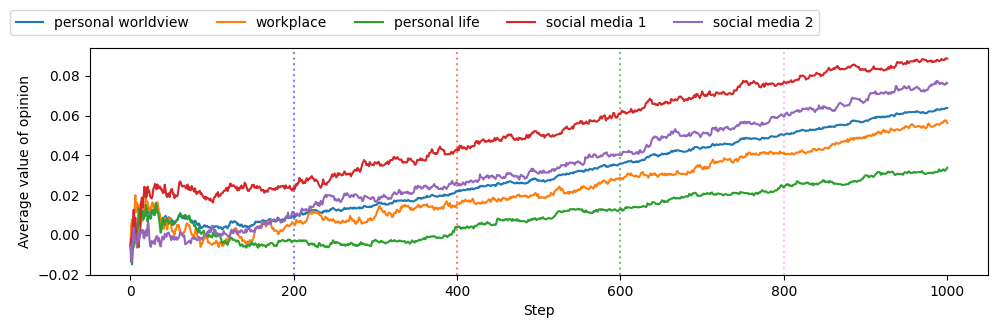}
\caption{A graph illustrating changes in average values of opinions for each of the multiplex network layers.} \label{fig:exp1_avg_graph}
\end{figure}

\par
The initial random distribution of the opinions was randomised - causing the average value of opinion (which can be any float value between -1.0 and 1.0) to oscillate close to the value of zero. However, as can be observed in  Figure~\ref{fig:exp1_avg_graph}, these values began to change quite drastically in the initial phases of the experiment. Before the 200th step of the simulation, all except one of the social media layers seem to converge at zero. 
\par
In the later phases of the experiment, opinions appear to very slowly and steadily be rising, it is easy to predict that this trend will continue until all of the opinion layers would finally obtain one, common value after which point, no more modifications to one's opinion would be possible, as all of his or her neighbours from the given layer would be of the same mind on the given issue and no amount of interactions could further alter opinion's value. Discrepancies between layers, visible in this graph may be attributed to a topological separateness of some agents group, who in random process at the beginning might have been somewhat isolated from the wider population and because of that skew the overall value of the layer's average opinion in either direction. 
The line denoting the value of the personal opinion remains in the middle of the graph lines, as its value is always assigned as a simple average of the agent's opinions from other layers. 



\subsection{Conformism and other factors}
\subsubsection{Introduction.}
\par
This run of the experiment utilises all of the parameters described before. Therefore formula responsible for calculating how one's opinion will be affected (marked as $O_{al_{new}}$) by the interaction between two agents is more complex. When agents A and B interact, depending on how their parameters rank against each other, one of the four results may take place. In the Equation \ref{eq:opinion_change_formula_exp2} each of the cases is reliant on the values of $A_{b}$(authority of the agent B), $\bar{A}_{Nl}$ (average authority of all of the agent A's neighbours), $O_{al}$ (agent A's opinion in the layer l), $O_{bl}$ (agent B's opinion in the layer l) and  $\bar{\theta }_{Nl}$ (average openness of all of the agent A's neighbours).
\par
The first scenario is when agent B with whom interaction takes place is more authoritative than the average member of the neighbourhood of agent A but the difference between the opinions of A and B is larger than the average acceptance of agent A's neighbourhood. In this scenario even though the interlocutor appears to be very credulous, their opinion is too radical for the liking of agent A's environment, therefore model will choose to change agent A's opinion to the value of $ V(\bar{O}_{Nl},O_{bl})$, where $V$ is a function as described in Equation~\ref{eq:eq_opinion_change_exp1}.
\par
The second scenario is a straightforward situation in which individual A is interacting with an agent whose credentials are both of high authority and whose opinions are not too different from the other opinions in the neighbourhood. This will cause the agent to simply adopt this opinion as their own in this layer.
\par
The third case denotes a situation in which another individual is both untrustworthy when it comes to their authority, as  well as radically different in opinion from the rest of one's neighbourhood. This denotes a situation in which agent A would find it desirable to stray even further away from agent B's opinion and therefore, with the use of the $V$ function, would move toward the value of either 1.0 or -1.0, depending on which of these will move them further away from agent B's opinion. This kind of interaction may potentially cause some agents to paradoxically stray further away from the average value of their neighbourhood's opinion in what could be called a radicalisation process.
\par
The fourth scenario will describe a situation in which agent B's opinion is not too different from the neighbourhood's average, however, agent B is not too authoritative him or herself in which case agent A will simply consolidate their opinion with the rest of the neighbourhood by using the $V$ function.
\begin{equation}\label{eq:opinion_change_formula_exp2}
        O_{al_{new}}=
\left\{ 
  \begin{array}{ c l }
    V(\bar{O}_{Nl},O_{bl}) & \quad \textrm{,if } A_{b} > \bar{A}_{Nl} \textrm{ and } \left | O_{bl} - O_{al} \right | > \bar{\theta }_{Nl}\\

    O_{bl} & \quad \textrm{,if } A_{b} > \bar{A}_{Nl} \textrm{ and } \left | O_{bl} - O_{al} \right | < \bar{\theta }_{Nl}\\

V(O_{al},1\times sgn(O_{al}-O_{bl}))& \quad \textrm{,if } A_{b} < \bar{A}_{Nl} \textrm{ and } \left | O_{bl} - O_{al} \right | > \bar{\theta }_{Nl}\\

V(O_{al},\bar{O}_{Nl}) & \quad \textrm{,if } A_{b} < \bar{A}_{Nl} \textrm{ and } \left | O_{bl} - O_{al} \right | < \bar{\theta }_{Nl}\\
  \end{array}
\right.
\end{equation}

\subsubsection{Overview of the experiment}
\par 
The experiment has taken place over $n=1000$ steps and included a population of $1000$ agents. Similarly to the previous one, all of the values have been initialised to random values from within the scope of each consecutive coefficient at the beginning of the simulation.
\par
\subsubsection{Results of the experiment.} 
\begin{figure}
\includegraphics[width=\textwidth]{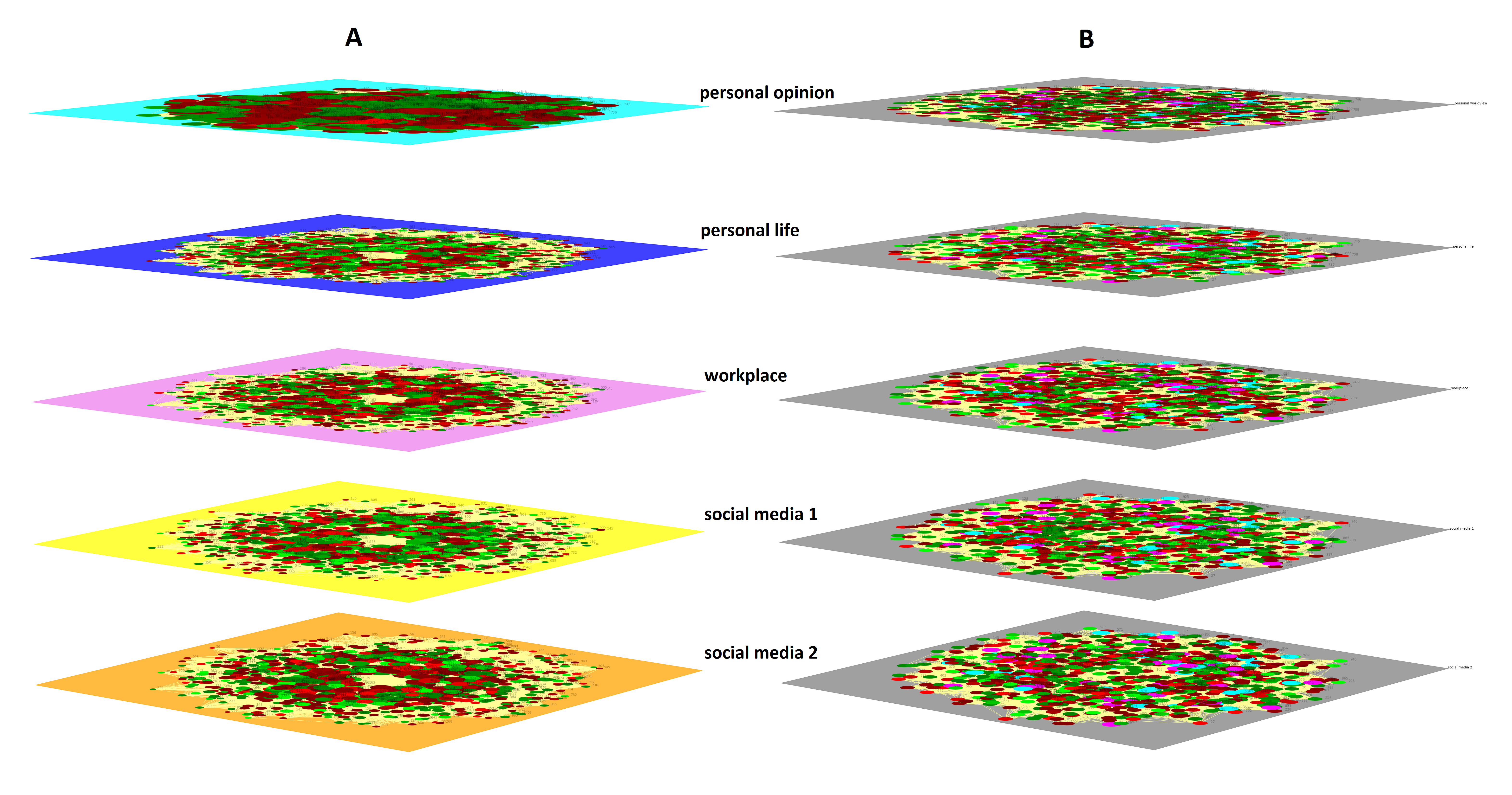}
\caption{Graphical representation of the multiplex network at the beginning of the simulation (network A) and after the experiment (network B).} \label{fig:exp2_before_after}
\end{figure}

Throughout the run, most of the more radical opinions held by the agents have been somewhat moderated which is perhaps best visible in the comparison between network 1 - a network at the beginning of the experiment and network 2 - network after \textbf{1000} steps of the simulation in Figure~\ref{fig:exp2_before_after}. Bright green and red colour marks these opinions which are closer to either \textbf{1.0} or \textbf{-1.0} and therefore are more radical, whereas darker shades describe opinions which are closer to the value of \textbf{0.0}. As one may easily notice, the second image is mostly devoid of these brighter spots. Nevertheless, the distribution of red and green colours shows only subtle changes (mostly in the 'Personal opinion' layer) between these two images. Moreover, in the image on the right side, nodes marked with the magenta colour indicate agents whose opinions have shifted by at least 1.0 downwards throughout the duration of the experiment and with the cyan colour those, which , similarly, have shifted significantly upwards.

\begin{figure}
\includegraphics[width=\textwidth]{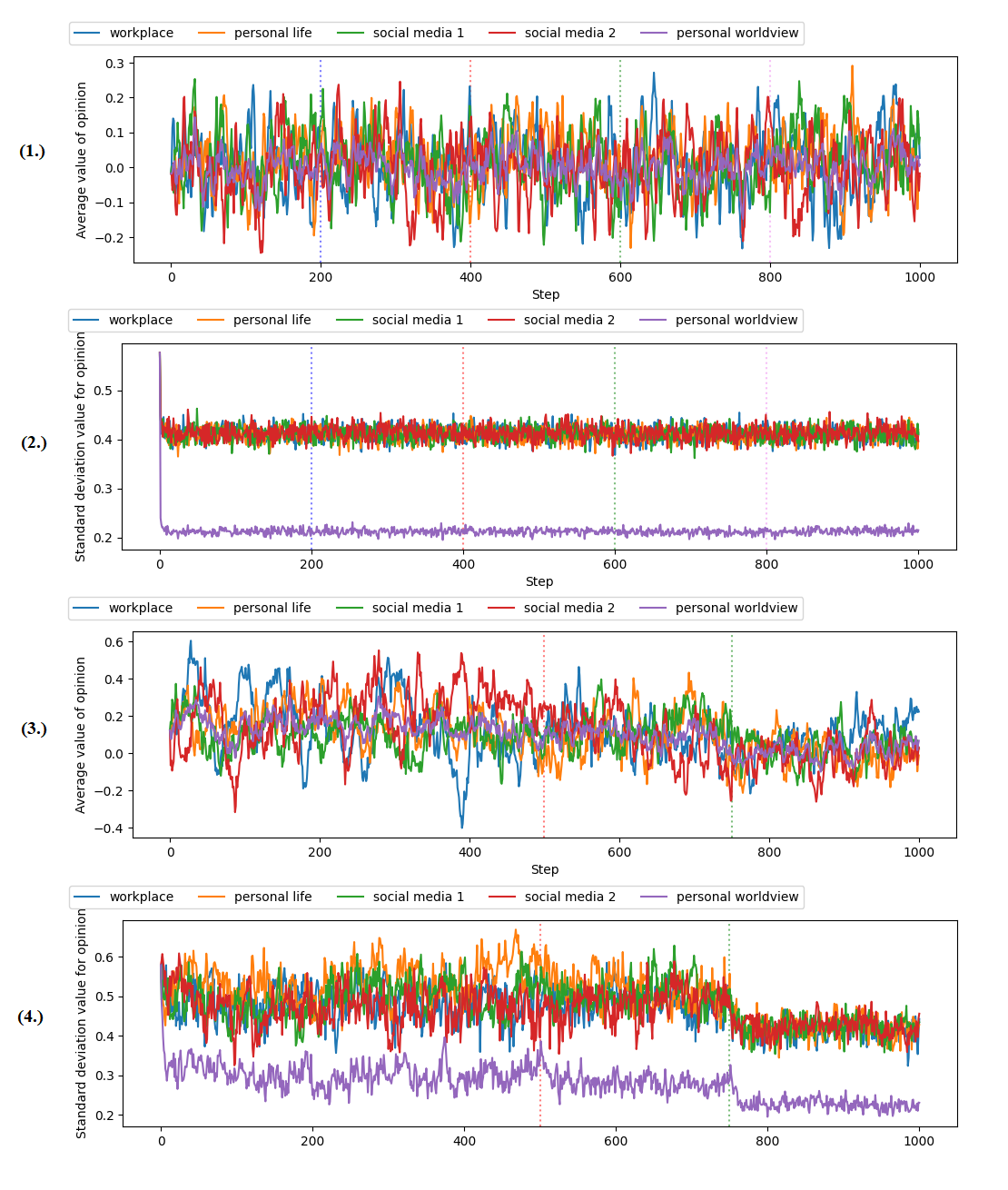}
\caption{From the top: \textbf{1)} A graph illustrating changes in the average value of opinions in the consecutive layers throughout the run of the second experiment. \textbf{2)} Graph showing values of the standard deviation of opinions throughout the second experiment. \textbf{3)} Graph of the average value of opinion throughout the run of the simulation with the injections of new agents during the run of the experiment. \textbf{4)} Graph of the measure of the standard deviation of agents' opinions during the run of the experiment with the injections of new agents. } \label{fig:exp2_std_avg}
\end{figure}

\par
As the result of the effect new coefficients bring into the model, as it is visible in Figure~\ref{fig:exp2_std_avg}.1, the average measure of opinions tended to fluctuate quite intensely when compared with the first experiment in which their values appeared to slowly rise, with minimal oscillation (as illustrated in Figure \ref{fig:exp1_avg_graph}). There is no trend towards convergence within any of the network layers. Since every interaction between any two neighbours occurs randomly (the likeliness of it taking place is predicated upon one's sociability parameter), more visible fluctuations were due to relatively rare occurrences in which a comparatively remote or unsociable agent or a group of agents happened to interact with the rest of the group and therefore skewing their average opinion. Worth noting is the fact that almost all of such fluctuations came to be as a result of more than one of such unlikely interactions (as may be deduced from the multi-step shape of their slopes). These were however quite rare as compared with the others and none lead to a new, long-standing trend in the shape of the graph.

\par
The stability of opinions in this approach, as compared to a simpler model from Experiment 1, may also be observed in Figure~\ref{fig:exp2_std_avg}.2. Here, the standard deviation of actors' opinions has reliably remained under the value of \textbf{0.5} for the entirety of the experiment. This lack of volatility is also reminiscent of the first experiment's results where the model's values did not take long to converge and settle on a value close to zero to synchronously drift towards one later on. What is different here is how large the value of convergence remains as compared to the previous simulation. This is an effect of how new mechanics take into the account parameters such as authority and openness to new views. These cause the system to form some sub-groups of agents who refuse to have their opinion changed by the current set of their neighbours in any further capacity and unless a new neighbour with perhaps a bigger authority or views a little bit closer to their own is introduced into their neighbourhood.

\par
To elaborate on the idea of altering this balance of opinions, a set of simulations was run which included injections of new agents into the system while the experiment was still in progress. What at first has been a small population of \textbf{25\%} of the original size, was then injected at step \textbf{500} with another \textbf{25\%} of this number and then, at step \textbf{750}, again with the remaining \textbf{50\%}. This could be interpreted as a real-life occurrence where for example a large new population of individuals is introduced to the local community as a result of, e.g. a merger between two companies which caused people to work with a larger number of co-workers. Here however this injection of new individuals took place in each of the layers simultaneously and, similarly as before, all the links between individuals were predicated on one's sociability. In Figure~\ref{fig:exp2_std_avg}.3 it is well visible how the average value of opinion has been affected after each of the consecutive injections of new agents into the simulation. They both caused the value of the opinions to become less spread out. Initial averages were oscillating quite violently at first but after the introduction of the remainder of the population, they lost some of that volatility and returned to the state of greater balance, with values more akin to these in Figure~\ref{fig:exp2_std_avg}.1.

\par
Figure \ref{fig:exp2_std_avg}.4 illustrates this trend even more clearly, as values of the standard deviation of opinions in each layer gradually converge towards lower value as new agents are being added into the population. With every new set injected into the simulation, the distribution of opinions agglomerates at a lower magnitude.

\section{Conclusions}
\par

In this work, we proposed a multilayer model of continuous opinion dynamics that incorporates multiple factors known from the literature, such as conformism, sociability or authority.. By conducting the simulations, we showed that despite the possibility of divergence of one's opinions at different layers, the network is capable of converging eventually in a model with only conformism implemented. However, if we let other factors play a role too, it is clearly visible that the opinions do not converge on all layers, but if new actors appear over time, they are capable of reducing the differences. This demonstrates that for the latter case, none of the aspects is strong enough to dominate others. The code for the model has ben released as the Code Ocean capsule\footnote{The code of the model: \href{https://doi.org/10.24433/CO.1006113.v1}{https://doi.org/10.24433/CO.1006113.v1}}.

The extensions of this work will focus on investigating the interplay between all components in more detail to answer the question of which of these can be thought of as the most important in disturbing the lack of consensus.

\subsubsection{Acknowledgements} This work was partially supported by the Polish National Science Centre, under grants no. 2021/41/B/HS6/02798~(S.S. \& J.J.), 2016/21/D/ST6\-/02408~(R.M.) and 2022/45/B/ST6/04145~(P.B.). This work was also partially funded by the European Union under the Horizon Europe grant OMINO (grant no. 101086321). Views and opinions expressed are however those of the author(s) only and do not necessarily reflect those of the European Union or the European Research Executive Agency. Neither the European Union nor European Research Executive Agency can be held responsible for them.
%
%
%
\bibliographystyle{splncs04}
\bibliography{bibliography}
%




\end{document}